Pierre Depaz

# Critiques protocolaires d'Internet: comparaison des projets IPFS et SecureScuttleButt

**Pierre DEPAZ**

pierre.depaz@sorbonne-nouvelle.fr

# Critiques protocolaires d'Internet: comparaison des projets IPFS et SecureScuttleButt

Pierre Depaz est un doctorant à l'ED120 Paris-3 Sorbonne Nouvelle. Sa thèse, sous la direction d'Alexandre Gefen (Paris-3) et Nick Montfort (MIT) porte sur la relation entre esthétique et compréhension au sein des codes sources, et sa recherche s'intéresse plus largement sur les pratiques sociales médiatisées par des systèmes techniques.

*Cet article explore deux propositions d'infrastructures critiques comme alternatives à l'état actuel des protocoles de l'Internet: IPFS (Interplanetary File System) et Scuttlebutt, mettant en avant les a priori et débats politiques de ces entreprises techniques. Pour ce faire, je propose d'analyser les discours des développeurs et développeuses de ces deux systèmes sur le mode d'une analyse de discours critique.*

*Cet article met en avant une forme de critique particulière aux régimes d'Internet: la critique infrastructurelle, et en souligne la variété par une étude comparative. À travers ces deux cas d'étude, nous verrons comment différentes alternatives aux actuelles implémentations spatio-temporelles d'Internet permettent d'identifier les dimensions d'agencement de ces actes de détournements et de substitution, caractérisant deux approches bien différentes de protocoles décentralisés, pourtant liés par une similarité technique.*

*Mots-clés: études des médias, protocole, analyse comparative, infrastructure.*

*This paper explores two critical infrastructure proposals as alternatives to the current state of the Internet protocols: IPFS (Interplanetary File System) and Scuttlebutt, highlighting the political a priori and debates of these technical enterprises. To do so, I propose to analyze the discourses of the developers of these two systems in the mode of a critical discourse analysis.*

*This article highlights a particular form of criticism of Internet regimes: infrastructural criticism, and highlights its variety through*

Pierre Depaz

*a comparative study. Through these two case studies, we will see how different alternatives to the current spatio-temporal implementations of the Internet allow us to identify the agency dimensions of these acts of hijacking and substitution, characterizing two quite different approaches to decentralized protocols, yet linked by a technical similarity.*

*Keywords: media studies, protocol, comparative analysis, infrastructure.*

*Este artículo explora dos propuestas de infraestructuras críticas como alternativas al estado actual de los protocolos de Internet: IPFS (Interplanetary File System) y Scuttlebutt, destacando las ideas preconcebidas y los debates políticos de estas aventuras técnicas. Para ello, propongo analizar los discursos de los promotores de estos dos sistemas al modo de un análisis crítico del discurso.*

*Este artículo destaca una forma de crítica particular de los regímenes de Internet: la crítica infraestructural, y pone de relieve su variedad mediante un estudio comparativo. A través de estos dos estudios de casos, veremos cómo diferentes alternativas a las actuales implementaciones espacio-temporales de Internet nos permiten identificar las dimensiones de agencia de estos actos de secuestro y sustitución, caracterizando dos enfoques bastante diferentes de los protocolos descentralizados, aunque vinculados por una similitud técnica.*

*Palabras clave: estudios de medios de comunicación, protocolo, análisis comparativo, infraestructura.*

Critiques protocolaires d'Internet: comparaison des projets IPFS et SecureScuttleButt

## 1. Introduction

Avec toutes leurs implications économiques, sociales et politiques, l'Internet et le Web sont avant tout des protocoles de communication, c'est-à-dire un ensemble de règles permettant à deux parties ou plus de requérir et fournir des données dans un même réseau. Cette suite de protocoles voit d'abord le jour autour de TCP/IP à la fin des années 1960 sous l'égide de la recherche militaire étasunienne, tandis que les deux protocoles HTTP et HTTPS qui constituent l'infrastructure du Web sont développés et distribués par le Centre Européen de la Recherche Nucléaire, une institution publique de recherche. Ces deux documents sont mis-à-jour, avec la version 6 de l'Internet Protocol et la version 3 de l'HyperText Protocol étant actuellement (2022) en cours d'adoption.

Pourtant, l'implémentation et l'adoption de ces protocoles ont découlé sur des utilisations bien différentes de leurs usages intialement envisagés—i.e. le partage de fichiers textes entre centres de recherches. Cette évolution est notamment documentée par Lawrence Lessig, dans son ouvrage *Code and Other Laws of Cyberspace* (Lessig, 1999), en ce qu'il identifie différentes forces ayant faconné l'évolution de l'Internet et du Web: des forces légales, marchandes, sociales et technologiques[1]. Par example, Harsh Gupta s'interroge sur le manque de représentation des contients africains, sud-américains et asiatiques lors des délibérations ayant pour objet l'implémentation de l'Encrypted Media Extensions (EME) (H. Gupta, 2016). Dans cet example précis, l'EME est un standard de communication pour contenus protégés par une propriété intellectuelle, une propriété intellectuelle de tradition majoritairement occidentale, et désormais établie en tant que vérité technique plutôt que réglementation économico-politique. Dans ce cas-là, il semble que le protocole en lui-même comporte une capacité d'influence et de détermination du comportement de l'utilisateur.

Les dérives de surveillance, de limitation de partage, d'automatisation du comportement et de monopole des applications issues des protocoles Internet et Web sont donc bien documentées (Lovink, 2011). Face à celles-ci s'élèvent alors plusieurs types de critiques: critiques sémantiques, sous la forme de blogs, de livres, d'articles et de conférences; critiques légales, telles que les licences GPL ou Creative Commons (Elkin-Koren, 2006) ou les législations de la RGPD ou du DSA; ou encore critiques programmatiques, telles que les bloqueurs de publicités (R. Gupta & Panda, 2020) ou des applications liées à des performances politiques (Critical Art Ensemble, 1996).

---
[1] Des analyses notamment confirmées par Dominique Cardon dans son ouvrage *Culture Numérique* (Cardon, 2019).

Ces différentes critiques sont donc toutes des manières d'exposer autant les limitations que les alternatives à un objet donné à un moment donné, se focalisant souvent sur un ou plusieurs points contentieux. La critique sémantique est rhétorique, et offre des stratégies discursives, la critique légale déploie un appareil d'arguments valides en termes législatifs, et la critique programmatique promeut l'utilisation de de dispositifs d'actions (dont des logiciels tels que deCSS, uBlock Origin, Consent-O-matic font partie) pour remédier aux limitations identifiées de manière pratique et computationelle. Le type de critique sur lequel je vais me pencher ici est celui de la *critique protocolaire*.

Un protocole est donc un système décrivant comment plusieurs parties peuvent échanger de l'information, en spécifiant une syntaxe, une sémantique et une synchronisation des messages échangés. Un protocole est ensuite un implémenté, soit par des logiciels, ou par du matériel informatique. Partant du principe, selon Alexander Galloway, qu'un protocole encode des manières de faires qui contraignent ses utilisateurs à la suivre sous peine d'être exclus de la communication se déroulant à travers ce protocole (Galloway, 2004), j'envisage ici la critique protocolaire comme la conception et la distribution d'infrastructures abstraites—complémentées par des implémentations concrètes—qui addressent les limitations identifiées d'une infrastructure existante.

Une critique protocolaire est donc une proposition alternative, autant discursive qu'opérationelle, aux défauts perçus d'une manière existante de communiquer. Nous considérons qu'elle opère en trois temps. Premièrement, elle met à jour les limites supposées du ou des protocoles existants, indiquant des défauts quant à la relation entre description théorique et utilisation pratique de ces protocoles, et proposant une approche alternative. Deuxièment, elle propose une alternative technique remédiant aux limites décrites en premier lieux, décrivant une manière de faire différente; celle-ci est donc opérationelle, et sa validité comme critique dépend en partie sur son utilisation pratique. Enfin, le troisième partie de la critique protocolaire découle de cet aspect pratique, en ce qu'elle se base sur l'utilisation effective et l'existence d'applications concrètes du protocole. Une des composantes centrales de cette critique protocolaire est donc un

Critiques protocolaires d'Internet: comparaison des projets IPFS et SecureScuttleButt

certain déterminisme technologique: changer une manière de faire peut changer un système socio-informationnel.

De cette possibilité de critique protocolaire découlent plusieurs questions que nous aborderons à travers la comparaison de deux études de cas: celle du protocole IPFS (Interplanetary Filesystem) et celle du protocole SSB (Secure Scuttlebutt). Ces deux protocoles se basent tous les deux sur une technique dite de liste chaînée, qui permet d'assurer une sorte de permanence de l'information en inscrivant l'unicité d'un contenu dans une continuation de contenus existants; ils trouvent également leur origine dans des limitations du protocole dominant d'Internet d'HTTP. En revanche, nous verrons que les conceptions dominantes du genre d'information qui doit être garanti (objet ou personne) diffère, de même que les manières de présenter l'influence d'un protocole technique sur des postures socio-culturelles.

Il s'agira d'examiner, dans les deux cas, les capacités expressives des protocoles numériques en tant que sous-ensemble des systèmes computationnels, en se basant notamment sur les travaux d'Ian Bogost en rhétorique procédurelle, ainsi que les possibilités de déterminisme technologique en comparant les usages abstraits imaginés par le protocole avec ses implémentations concrètes par des acteurs situés (Bogost, 2007). Tandis que Bogost se focalise sur les discours émanants de systèmes interactifs pour l'utilisateur final, nous nous concentrons ici sur ces discours suggérés par une infrastructure de communication, et donc relativement invisibles à l'utilisateur final. Ces infrastructures de communication tendant à une base massive d'utilisateurs, nous nous appuyons également sur les travaux de Francesca Musiani, et notamment son analyse des compromis apparaissant entre promesse décentralisée et réalité centralisée (Musiani, 2017).

Cela nous amènera enfin à considérer à quel point ces protocoles tentent de proposer des nouveaux imaginaires possibles en ce qu'il s'agit de considérer l'échange d'information sur des réseaux numériques, et plus spécifiquement quant aux façons de considérer, techniquement, l'objet, l'individu, l'espace et le temps, et comment ces imaginations se réalisent en pratique. Ainsi, cet article s'inscrit dans le champ des STS (études des sciences et techniques), et plus particulièrement dans les champ des *software studies*, cherchant à expliciter les effets et influences socio-culturellles des logiciels (Fuller, 2003).

Pierre Depaz

    Comment se constitue en pratique une critique protocolaire? Quelle est la place de la technique-même au sein de cette critique? Comment les environnements, documents et actions sociales, économiques et techniques constituant un protocole reflètent-ils des conceptions plus larges de manières de communiquer? Afin d'élucider ces questions, nous procéderons à une analyse du protocole et du discours des deux écosystèmes d'IPFS et SSB, confrontant language machine et language humain. Cette analyse consistera à identifier les éléments techniques et discursifs décrivant un protocole: le normatif (le protocole en lui-même), le prescriptif (les usages imaginés par les concepteurs), le descriptif (la représentation du projet à travers sites webs, entretiens dans la presse et promotion individuelle) ou encore l'argumentatif et le participatif (discussions entre concepteurs et utilisateurs autour des intentions et usages des protocoles). Cette analyse du protocole inclut donc le technique et l'humain.

    Le cadre d'interprétation de ces documents sera ici celui d'une analyse critique du discours, telle qu'elle est développée par Dianna Mullet, partant de l'hypothèse que les différents facettes du discours d'une même organisation permettent alors de mettre à jour une certaine cosmogonie suggérée avec, à sa base, un protocole comme élément socio-technique (Mullet, 2018). Nous considérons que le choix de cette méthodologie d'analyse de discours est ici approprié en ce que nos objets d'études sont des objets communicationnels multiformes, entrelaçant des discours techniques, des discours politiques et des discours sociaux, aussi bien dirigés vers une communauté interne que vers un plus grand public. L'analyse des présupposés, des champs lexicaux et des addresses des auteurs et autrices de ces discours, considérés dans la dimension pragmatique de l'adoption des ces protocoles, nous permettra de constituer les tenants et les aboutissants de la critique protocolaire déployée par IPFS et SSB.

    Cette approche d'analyse critique du discours a donc lieu au sein d'une analyse comparative, et cela pour deux raisons. Premièrement, il s'agit de mettre en exergue les éléments communs au déploiement d'un protocole: pendant technique, pendant communicationel, et pendant social, et de voir comment la manifestation de ces éléments peut varier selon les présupposés des concepteurs, alors qu'ils restent très semblables au niveau technique. Deuxièmement, il s'agit de considérer l'implication d'un même but (communication d'un message d'un émetteur à un récepteur), avec un même principe technique (chaîne d'information

cryptographiée) dans des environnements sociaux différents. Nous analyserons donc à quel point ce but et ce moyen technique résultent, ou non, en des conséquences pratiques différentes, notamment en termes de type et de nombres d'applications développées.

Nous approcherons ce sujet en deux temps. Tout d'abord, nous examinerons les visions paradigmatiques des deux projets, en commencant par IPFS, suivi de SSB. Ces examinations se feront de manière identique, à travers la description du protocole, puis l'identification des discours mis en place autour de celui-ci, pour enfin conclure sur la mise en pratique et les possibles limites de la confrontation au réel que chaque protocole présente avec travers ses applications. Ensuite, nous approfondirons notre comparaison en considérant la composante du déterminisme et des ontologies telles qu'elles se manifestent dans chaque protocole.

## 2. IPFS: une disponibilité permanente de l'objet

### 2.1. Description du protocole

L'*InterPlanetary File System*, ou système de fichiers interplanétaires, est un protocole de distribution d'information qui considère comme primordial la disponibilité permanente de tout objet, et propose donc une organisation technique dans cette optique. La première version du protocole est spécifiée en 2014 par Juan Benet, sous la forme d'un *white paper*, publication scientifique non revue par les pairs, mais néanmoins décrivant un problème particulier, et certaines solutions envisagées (Benet, 2014). Ce document est fondateur de l'approche d'IPFS, et révèle que c'est un projet lui-même constitué d'un amalgames de protocoles existants, composés afin de réguler la création et la vérification d'identité de chaque pair du réseau, la connection et la localisation de pairs au sein du réseau, l'échange d'information entre pairs, et surtout la représentation des objets hébergés par les pairs, intégrant étroitement les questions de versions et de nomination, dans ce qui est considéré un des piliers de l'approche IPFS, le *content-addressing*[2]. Bien que sa nature décentralisée rend difficile d'estimer l'étendue exacte de son utilisation, le réseau

---

[2] Ce *content-addressing* est une manière de faire référence à une entité, non pas par son addresse de sa localisation (comme pour une URL classique), mais par la nature de son contenu-même (e.g. suite de pixels, de caractères)

Pierre Depaz

IPFS était constitué d'environ 40,000 noeuds en 2020 (Henningsen et al., 2020).

Critiques protocolaires d'Internet: comparaison des projets IPFS et SecureScuttleButt

IPFS recombine des technologies existantes, puisant depuis divers domaines de la cryptographie et de la théorie des graphes, pour permettre la disponibilité à tout un chacun d'un seul ensemble globalisé de fichiers. D'après les termes de Juan Benet,

> *IPFS est similaire au Web, mais peut être vu comme un seul essaim BitTorrent, échangeant des objets au sein d'un dépôt Git. En d'autres termes, IPFS fournit un modèle de stockage en bloc à haut débit, avec des hyperliens adressés au contenu. Cela forme un Merkle DAG généralisé, une structure de données sur laquelle on peut construire des systèmes de fichiers versionnés, des blockchains et même un Web Permanent.[3] (Benet, 2014)*

Ce que nous notons *a priori*, c'est donc un discours qui fait la place belle à cet enchevêtrement de technologies, se présentant comme similaire au Web actuel, mais aussi se présentant comme une version améliorée, car permanente. Le Web est en effet considéré par Benet comme éphémère et temporel, l'antithèse d'IPFS—le web permanent et l'accessibilité universelle de toute information[4]. Parmi ces technologies figurent une DHT, *Distributed Hash Table*, un système d'incitation de partage BitSwap, et une représentation d'objets à travers un graphe acyclique de Merkle, eux-mêmes accessibles à travers une infrastructure de clés publiques. Ces innovations techniques vont se combiner pour réaliser la vision d'un protocole assurant un partage de l'information supposé global et permanent, intégrés au système de fichiers "normal" de l'utilisateur.

---

[3] "*IPFS is similar to the Web, but IPFS could be seen as a single BitTorrent swarm, exchanging objects within one Git repository. In other words, IPFS provides a high through-put content-addressed block storage model, with content-addressed hyper links. This forms a generalized Merkle DAG, a data structure upon which one can build versioned file systems, blockchains, and even a Permanent Web*"

[4] *Un web où [...] la publication d'informations précieuses n'impose pas leur hébergement à l'éditeur mais aux personnes intéressées, où les utilisateurs peuvent faire confiance au contenu qu'ils reçoivent sans faire confiance aux pairs qui le leur transmettent, et où les fichiers anciens mais importants ne disparaissent pas.* (Benet, 2014)

Pierre Depaz

   L'infrastructure de clé publique permet avant tout d'adresser les objets présents au sein du réseau par leur *contenu* plutôt que par leur *addresse*, un choix sur lequel nous reviendrons, afin de pallier à la disparition du contenu à l'adresse spécifiée—un phénomène manifesté sous la forme de la familière erreur 404 par le protocole HTTP. Les identifiants uniques de chaque objet sont ensuite répertoriés, de manière distributive, sur cette DHT, de sorte à ce que chacun des membres du réseau héberge la liste des objets disponibles, au moyen d'une liste chaînée. Cette façon de faire se pose en réponse au système DNS d'Internet, qui fonctionne lui de manière relativement plus centralisée[5]. Le transfert des objets se fait ensuite par le méchanisme BitSwap, qui récompense les membres du réseaux partageant le plus de contenus, et pénalisant ceux qui ne le font pas, à travers un système de dettes et de crédits. Enfin, la représentation de ces objets (qui peuvent être un fragment de texte, un fichier MP3, une section d'image, etc.) est techniquement *immuable*. Cela signifie que chaque objet, une fois inscrit au sein du réseau, ne peut être supprimé par son auteur, et ne disparait qu'une fois que tous les membres du réseaux ont cessé de l'héberger. Si une version subséquente de cet objet est ajoutée sur le réseau, il s'agit d'un tout nouvel objet, avec une nouvelle addresse, et existe donc en supplément de l'objet précédent.

   Benet présente donc cette permanence, c'est-à-dire la disponibilité de chaque objet à n'importe quel membre du réseau, aussi longtemps que ces membres décident de le conserver, comme étant rendue possible par des techniques liant unicité de l'objet et mécanismes économiques de partage basé sur des dynamiques de marché, afin de pallier à ces myriades d'erreurs 404 du Web contemporain.

   Le premier tenant de la critique protocolaire déployée par IPFS se focalise donc sur le caractère éphémère du protocole HTTP, tel qu'il finit par impliquer une dépendance à une architecture client-serveur. Face à cette limite est tout d'abord présentée une solution technique, se basant sur notamment sur le concept d'une liste chaînée. Cette critique va ensuite être étayée par des discours des créateurs d'IPFS et des applications de leurs utilisateurs.

---

[5] Bien qu'également distribué, DNS présente des aspects centralisés et hiérarchiques, notamment par la présence de serveurs de noms faisant authorité et les serveurs de nom racine (e.g. l'attribution d'un nom racine *.com* ou *.fr* dépend en pratique d'une entité légale spécifique).

### 2.2. IPFS: vision du monde et réalité

Les raisons pour lesquels IPFS cherche à garantir un accès universel et atemporel à tout utilisateur est basée sur une critique de l'état actuel de l'Internet, et plus particulièrement du concept d'architecture centralisée, telle que Benet l'explique dans une de ses premières interventions publiques en 2016 à Stanford (stanfordonline, 2015), puis en 2019 à IPFS Camp. Il y a, au sein de la manière dont IPFS se présente, une dimension téléologique indéniable: lors de la première conférence IPFS, il inscrit IPFS dans la directe lignée des 10 millions d'années d'existence de l'espèce humaine[6]. Pour compléter cette approche d'une évolution linéaire, il pose alors le Web actuel comme inefficient en termes de coût par bande passante, du fait de son architecture client/serveur centralisée. C'est bien cette architecture centralisée qui est considérée par Benet et Protocol Labs comme rendant un document éphémère dès lors que ce dernier n'est plus hébergé par le serveur le fournissant. Enfin, il pose le développement économique de l'Internet et du Web aujourd'hui comme tendant à une centralisation et un monopole de l'accès à l'information qui sont considérées comme un obstacle à l'innovation.

Déjà, nous voyons que certaines sont des critiques technologiques concrètes (i.e. questions de bande-passante et de pérennité du contenu en ligne), mais les deux dernières sont plus floues, et plus difficilement attribuables à une technologie plutôt qu'à un ensemble de décisions socio-économiques, telles qu'identifiées par Lessig. Néanmoins, la vision du monde proposée par IPFS est celle d'un réseau de connexions perpétuel et quasi-instantané. Puisque chacun est considéré responsable à titre égal de la mise à disposition de l'information du réseau, IPFS doit permettre à chacun d'accéder à tout en permanence, là où l'Internet établit une relation de hiérarchie entre serveur et client.

Le discours d'IPFS établit donc une approche solidaire de la distribution d'information, faisant la part belle au contenu plutôt qu'à l'adresse de ce contenu. La réponse technologique apportée aux limitations d'Internet mentionnées plus haut—lenteur et disparition—est radicalement opposée. Il s'agit désormais de faire en sorte que chaque objet mis à disposition sur le réseau puisse y rester tant qu'*au moins un* pair décide de le fournir au réseau. Cette emphase revêt un caractère particulier lorsque Benet compare IPFS au rêve original du Web, par lequel Tim Berners-Lee imagine un réseaux de pairs. Et pourtant, nulle part dans les spécifications HTTP 1.0 et 1.1 figurent la mention de pairs, ou d'échange réciproque d'information. Cela semble alors être une sorte de révisionnisme historique, peut-être influencé par le projet plus récent du créateur du Web, Solid, lequel propose un moyen de décentraliser des applications sociales en se basant sur le concept de *Linked Data*.

---

[6] Présentant une vision déterministe de la technologie, notamment en considérant que la technologie de l'écriture a permis l'apparition de la culture.

IPFS addresse également le problème d'incitation à la distribution d'un contenu qui n'est ni celui que l'on possède, ni celui que l'on désire. Sous le régime protocolaire du Web, le serveur est toujours considéré comme ayant un intérêt à distribuer son propre contenu, tandis que le client sait qu'il s'addresse à un serveur spécifique afin de récupérer un contenu spécifique. Un réseau distribué doit, lui, se reposer sur le partage constant d'information, y compris une information qui n'est pas immédiatement pertinente aux utilisateurs les hébergeant—et donc sans incitation personelle intrinsèque. En universalisant les contenu, on court alors le risque de dépersonaliser le rapport au contenu et sa responsabilité.

Afin de pallier à cette limitation, IPFS propose BitSwap, une manière d'accumuler du crédit ou du débit en tant que réputation, au sein d'une logique de libre-échange (*marketplace*, selon les termes de la documentation (Protocol Labs, n.d.)), et dont l'auteur lui-même reconnaît dans le *white paper* qu'elle serait particulièrement adaptée à une cryptomonnaie. Celle-ci sera développée sous la forme d'un FileCoin au même moment qu'IPFS—la composante principale du partage de contenu est donc d'inspiration financière—un système de troc tel que BitTorrent ne fonctionne plus dès lors qu'il s'agit d'acquérir une multiplicité de fichiers, plutôt qu'un seul (Protocol Labs, 2014).

Le protocole IPFS vise ainsi à la permanence du contenu hébergé sur la plateforme indépendamment du ou de la propriétaire, et que ces créateurs considèrent que cette fonctionalité est intrinsèquement désirable, malgré les risques encourus par la permanence de contenus illégaux. Techniquement, une telle permanence peut être accomplie par le biais d'incitation financière à travers une cryptomonnaie, et non à travers la manière dont HTTP le faisait, c'est à dire la responsabilité individuelle de l'hébergeur, ou la manière de BitTorrent, en se focalisant sur un seul fichier, sur un traqueur d'information centralisé et sur la réciprocité de la distribution d'information. Ce système de stockage de l'information à travers FileCoin propose alors un espace de stockage relativement

Critiques protocolaires d'Internet: comparaison des projets IPFS et SecureScuttleButt

démesuré[7]. Si le protocole d'incitation marche, dans le sens où il y a bien un vaste espace de mémoire mis-à-disposition, c'est alors l'utilisation de ce protocole qui va nous intéresser—ce qui est fait de cet espace de mémoire. En effet, comme le dit Tony Willenber en 2016, dans sa présentation d'IPFS:

> *L'IPFS n'est pas une simple expérience théorique ou académique. Il s'agit d'un système logiciel fonctionnel (bien qu'encore en phase alpha) qui peut être téléchargé et mis en service dès maintenant.[8] (Willenberg, 2018)*

Un des aspects de la critique protocolaire est donc bien d'être opérationelle, c'est-à-dire de pouvoir se manifester directement dans des produits et des usages. En plus des discours sociaux présentant la vision originelle du protocole et les discours techniques assurant cette opérationalité, la cohérence ou discordance entre théorie et pratique se révèle dans les applications et cas d'usages.

### 2.3. Applications

La documentation du site IPFS propose une liste exhaustive de cas d'usages, aussi bien potentiels que déjà réalisés. On y retrouve notamment le partage de fichiers par un individu, la collaboration en temps-réel sur le même fichier ou encore l'utilisation comme messagerie. Cependant, la principale raison d'être d'IPFS est bien celle d'un protocole, c'est-à-dire en tant qu'infrastructure afin d'héberger, gérer et distribuer du contenu à travers le monde—par exemple, Netflix étudiait en 2021 la possibilité de synchroniser certains de ces systèmes à l'échelle globale via IPFS (Protocol Labs, 2020). Plus particulièrement, IPFS concentre son champ d'application de l'IPFS à celui des *dApps*, ou applications décentralisées traditionellement basées sur des systèmes de blockchain, ce qui annonce une certaine contingence de l'écosystème des blockchains avec celui d'IPFS.

Cette vision d'un monde informationel qui serait mieux si toute information était permanente se reflète dans plusieurs cas de figure mis en avant par Protocol Labs, la structure encadrant le développement d'IPFS. Dès 2014, Juan Benet présente le travail de l'organisation *Internet Archive* comme étant essentiel pour la sauvegarde des connaissances humaines à l'ère informatique, et

---

[7] 39 Petabytes pour un peu plus de 818000 objets distincts, soit un espace dédié moyen de 53 Gigabytes par objet. Données au 01/02/2022, https://storage.filecoin.io/

[8] *"The IPFS is not just a theoretical or academic experiment. It is a working software system (although still in alpha) that can be downloaded and switched on right now"*

Pierre Depaz

suggère donc d'utiliser IPFS afin de créer une copie des archives de l'*Internet Archive*. Un projet est donc ouvert sur GitHub où les développeurs et développeuses discutent de l'implémentation en 2015, avant qu'il soit abandonné en 2017, au même moment que se pose la question de "qui" va héberger cette archive— nominativement la communauté IPFS, mais en pratique cette responsabilité est endossée par les employés de Protocol Labs[9], retournant donc à un hébergement centralisé. En 2017, une autre application du protocole se déroule lors de la copie du site turc de Wikipedia sur IPFS alors que la version HTTP du même site est censurée par le gouvernment Erdoğan. Dans ce cas précis, il s'agit alors d'une initiative de Protocol Labs. Si leur annonce sur leur blog se présente comme étant pair-au-pair et décentralisée, l'organisation est, en pratique, la principale à héberger ce contenu, comme le montre leurs demandes subséquentes de co-hébergement par des tiers (The IPFS Team, 2017).

La question de la censure se pose à l'inverse pour le reste du contenu hébergé sur IPFS. En effet, le protocole se base sur l'immutabilité des contenus, ce qui mène donc à la conséquence de suppression de contenu illégal. Un problème compliqué et pourtant intrinsèque à toute information publique résumé sur le répositaire GitHub du projet par l'utilisateur *geebotron*, à propos de la modification des fichiers disponibles:

*Chaque fichier existant sur IPFS est susceptible d'offenser quelqu'un* [10] (geebotron, 2016).

En effet, la conception d'un protocole sur la permanence d'une information se heurte alors de manière frontale à la question de la censure et de la propriéte intellectuelle. Mis face à l'existence de législation regardant la propriété intellectuelle, la réponse d'IPFS est donc de rajouter un aspect technique additionel (i.e. *une whitelist*) pour déterminer l'utilisation des fichiers. La présentation de cette décision techniques est ensuite présentée comme clotûrant le débat:

---

[9] Voir notamment les dicussions sur https://github.com/ipfs-inactive/archives/issues/88: "*Ideally "the community" would be hosting it, but in practice I suspect a lot of stuff is only pinned on one of our nodes*", utilisateur davidar, 15.01.2017

[10] *"Every single file that could exist on IPFS has the potential to offend someone"*

Critiques protocolaires d'Internet: comparaison des projets IPFS et SecureScuttleButt

> *Je n'ai pas l'intention d'entamer une discussion de philosophie politique, mais plutôt d'articuler l'espace de conception et les raisons pour lesquelles IPFS s'inscrit dans un ensemble particulier de décisions. Si la question s'éloigne du sujet, je la fermerai ou la renommerai.*[11] (Benet, 2017)

Enfin, la question de l'applicabilité du protocole BitSwap se retrouve dans le développement de services d'entreprise centralisés[12]:

> Si vous n'épinglez pas votre contenu sur IPFS, il disparaît. Et si le serveur qui épingle vos ressources est mis hors ligne et que personne d'autre ne l'épingle, il disparaît à jamais. C'est pourquoi un marché s'est ouvert pour des services tels que Pinata, qui visent à être des services d'épinglage permanents. Vous vous retrouvez donc toujours avec un cadre commercial centralisé même si la technologie elle-même est décentralisée, en ce sens que si vous n'avez pas les moyens de fournir votre propre infrastructure distribuée, vous devrez payer quelqu'un qui le fera.[13]

En fin de compte, ce que nous voyons dans ce développement, c'est que la présentation par IPFS de leur critique protocolaire— limitations de protocole existant, aspect technique d'une alternative, discours justificatifs et applications—trouve une elle-même une limitation, puisque que centralisée au niveau économique plus qu'au niveau technique et se reposant notamment sur la responsabilité d'hébergement d'entités économiques (Protocol Labs, Piñata). L'insertion d'un facteur économique lors de l'application d'un protocole technique va donc contrecarrer un discours de décentralisation universelle: il semble alors que cette pluralité des composantes de la critique déjoue la conception de déterminisme technologique de Benet et Protocol Labs.

---

[11] "*It is not my intention to start a political philosophy discussion—but rather only to articulate the design space and why IPFS falls in a particular set of decisions. If the issue gets more off topic, i'll just close it, or rename it.*"

[12] Voir la discussions sur les applications d'IPFS: https://www.reddit.com/r/ipfs/comments/ruxlej/ipfs_is_an_alternative_for/

[13] *If you don't pin your content to IPFS, it goes bye-bye. And if the server pinning your resources ever goes offline and no one else has it pinned, it's gone forever. That's why a market has opened up for services like Pinata, which aim to be permanent pinning services. So you still end up with a centralized business framework even if the technology itself is decentralized, in that, if you don't have the means to provide your own distributed infrastructure, you're going to have to pay someone who does.*

Pierre Depaz

### 3. SSB: implémentation technologique d'un protocole social

#### 3.1. Description

SSB (ou Secure Scuttlebutt) est un protocole de communication créé par Dominic Tarr en 2014, la même année qu'IPFS. Alors que Juan Benet est un développeur issu de la Silicon Valley, Tarr est un navigateur néo-zélandais qui pose la disponibilité hors-ligne, ainsi que la nature sémantique des objets comme messages, en tant que fondation du protocole, sous le terme de *local first*. Les membres de l'organisation autour de SSB (le *Secure Scuttlebutt Consortium*) estiment à la fin de 2019 le nombre de pairs sur le réseau à au moins 10,000 (Fiscella, 2018).

SSB, un protocole de communication et de distributions de fichiers tout comme IPFS, va se démarquer tout d'abord par son positionnement dans la famille des *gossip protocols*, des protocoles de ragots. Ceux-ci sont basés sur des modélisations de la distribution des rumeurs, ou des épidémies—essentiellement une sélection plus ou moins au hasard des pairs à portée de communication auxquels l'information va se propager. Par exemple, dans le cas de Tribler[14], un protocole de partage de fichier sur le modèle de BitTorrent, il s'agit de diffuser de l'information en s'appuyant sur ses voisins immédiats, plutôt que sur une instance centrale. L'approche est donc posée: il s'agit de définir un protocole technique comme simulation de phénomènes naturels, partant du principe que l'information va avoir, en son sein, des schémas sociaux (Web3 Foundation, 2019).

SSB affiche comme axiome de départ la disponibilité hors-ligne: le protocole et ses applications doivent être utilisables lorsque l'on n'est pas connecté à l'Internet. Ainsi, tout stockage et accès de données se fait uniquement de manière locale, et la mise à jour, ou synchronization de ces données vont être effectués quand, et si, il y a une connexion effectuée à un ou une autre membre du réseau. SSB réplique, de manière protocolaire, l'expérience d'une vie en mer, expérience locale si il en est, en ne permettant la synchronisation entre deux pairs que si ceux-ci sont connectés au même réseau local.

SSB est également un protocole décentralisé de partage de données. Il fonctionne par pairs qui s'assignent une signature

---

[14] Voir la section "Why don't I see all torrents in a channel?" sur le Tribler Wiki: https://www.tribler.org/NewFAQ/#, consulté le 02.03.2022.

Critiques protocolaires d'Internet: comparaison des projets IPFS et SecureScuttleButt

cryptographique, afin de constituer leur identité—si on perd sa clé, on perd tout accès à son compte, corrélat de la responsabilité de l'utilisateur de ses propres données. La manière dont les pairs se découvrent, en revanche, ne se situe que lorsqu'ils émettent des signaux jusqu'à ce qu'ils soient reçus par un autre pair. Une fois que la connexion est établie entre deux pairs, ces derniers peuvent effectuer un échange de messages ou de fichiers. Il est également possible de se connecter à un pair à travers un portail Internet: SSB n'est donc pas directement une critique protocolaire cherchant à remplacer HTTP, mais plutôt à compléter son aspect technique[15]. La critique présentée par SSB est d'avantage dirigée vers l'injonction à être constamment "en-ligne", plutôt que considérer le "hors-ligne" comme approche par défaut.

Enfin, les messages eux-mêmes sont basés sur une conception da la liste chaînée, elle-même similaire au versionnage des fichiers d'IPFS. Nous retrouvons ici une similarité technique qui, mise face à la différence de discours et d'application, nous permet de questionner le présupposé du déterminisme technologique sous-jacent dans le fait que le point de départ de chaque critique protocolaire est une description technologique. Pour SSB, chaque message est relié cryptographiquement au message précédent et au message suivant, formant donc ce que les développeurs n'appellent pas une *chain*, mais un *feed* (flux). De cette manière, les objets ne sont plus uniquement considérés comme étant des entités flottantes, uniques, mais bien des entités relationelles qui tirent leur nature sémantique de leur contexte—un contexte avant tout social. Ici, la technique sous-jacente est la même, mais leur développement et présentation va suggérer des manières de communiquer bien différentes.

SSB met donc en avant à quel point leur protocole se base sur des phénomènes naturels, et présente la décentralisation non pas comme ayant une vocation à l'étendue universelle, mais bien à la responsabilité locale, manifesté dans le protocole même par une asynchronie de l'échange de messages, par une relation cryptographique, et par un stockage local, et donc personnel, de toutes données. Nous allons maintenant voir comment les discours autour de SSB renforcent cette préférence pour une approche sociale plutôt qu'une approche technique.

---

[15] Ceux-ci existent sous forme de *pub*, jeu de mot entre "bar" et "publiciste"

Pierre Depaz

### 3.2. Communication

Le discours de SSB, et notamment la manière dont est présenté le protocole, constitue un miroir intéressant d'IPFS: plutôt qu'une démonstration technique, il s'agit plutôt de prendre un cas d'étude —une relation amoureuse—pour montrer comment SSB supporte une conception intime de la communication, plutôt qu'une conception technique (Mandeville, n.d.-a). À travers une vidéo introductive, montrant comment SSB peut exister en arrière-plan d'une histoire romantique, se développe un discours bien différent de celui d'IPFS. La page de documentation du site de SSB, au lieu de se présenter sous la forme d'un article académique publié de manière autonome sur un site de recherche scientifique, accueille la visiteur avec le message suivant:

> *Scuttlebutt vise à harmoniser quatre perspectives de la vie : L'environnement reflétant la technologie reflétant la communauté reflétant la société.*
>
> *Nous reconnaissons les environnements naturel, virtuel et social. Notre responsabilité est de reconnaître quelles ressources sont abondantes, lesquelles sont suffisantes, et de nous adapter en conséquence grâce à l'efficacité.*[16]

Bien qu'il y ait également une possibilité de consulter une documentation plus poussée et plus rigoureusement technique, ainsi que l'article présentant le protocole SSB, publié dans les annales de la conférence Information-Centric Networking de l'ACM en 2019 (Tarr et al., 2019), il est intéressant de s'arrêter sur ce premier message d'accueil. Au vu de la place attribuée au terme *technologie*, coincé entre *environnement* et *communauté*, la position se présente comme étant bien plus relationelle que le *protocole hypermedia concu pour préserver et développer le savoir de l'humanité* d'IPFS, holistique et donc auto-suffisant.

Nous pouvons identifier cette prise de parole par une emphase sur la relation entre deux agents, plutôt que sur celle d'un réseau entier. Déjà, la première version du protocole SSB se base sur un article de Tarr considérant le problème d'authentification et

---

[16] "*Scuttlebutt aims to harmonize four perspectives of life: Environment reflecting Technology reflecting Community reflecting Society.*

*We acknowledge the natural, the virtual, and the social environments. Our responsibility is to recognize which resources are abundant, which are sufficient, and adapt accordingly through efficiency*"

Critiques protocolaires d'Internet: comparaison des projets IPFS et SecureScuttleButt

d'échange d'informations entre deux pairs[17] plutôt que sur la représentation d'une information. De cette manière, si ces deux protocoles, IPFS et SSB considèrent tous deux la mise en place d'un système de communication, IPFS se focalise sur ce qui est communiqué (messages/objets comme membres d'un graphe acyclique dirigé), tandis que SSB se focalise sur les individus voulant échanger un message (signature cryptographiques synchronisant leurs contenus).

En tant que critique, c'est un double présupposé fondamental d'un internet *contemporain* que SSB critique: l'idée que toute information, et donc tout membre du réseau qui héberge ou demande une information, doit être disponible en permanence et à l'échelle globale. Ce présupposé, considèrent Tarr et ses collaborateurs, a notamment des répercussions sur l'autonomie des membres du réseau par rapport à l'influence de monopoles économiques, et à laquelle ils veulent redonner une prépondérance (ce *local first* mentionné plus haut) par le biais d'une propagation par défaut sur un réseau local que par réseau global[18].

Comme le développe Zach Mandeville dans son essai, financé par SSB, *The Future Will be Technical*, on note également dans l'écosystème SSB une croyance en le développement des technologies pour améliorer le futur, mais également une prise en compte de la technique en tant qu'élément de la culture:

> *Les discussions sur le développement et les tutoriels sont une partie essentielle de notre communauté et ne doivent pas être occultées ou minimisées.*[19] *(Mandeville, n.d.-b)*

La technique est ici un élément fortement culturel et l'adhésion à un protocole est donc une adhésion à une vision partagée du monde, à un imaginaire collectif (Hall, 1997). Il semble alors que le champ d'application évolue aussi. De manière symétrique à la perception d'IPFS, qui veut un protocole qui concerne l'univers entier, SSB s'addresse au *Scuttleverse*. On peut considérer ce *Scuttleverse* comme un cosmos qui est mis au monde par le biais d'un lien technique basé sur un lien social—où, par exemple, les *groupes d'utilisateurs* s'appellent des *tribus*, rappelant les travaux de Yuk Hui sur la cosmotechnique, cet aspect des systèmes techniques existant au sein de cosmos différents (dans le sens de cultures différentes) (Hui, 2016). La cosmotechnique, selon Hui, suggèrent que différentes cultures ont différentes manières de penser leur relation à la technologie; ces différentes manières de penser impliquent ensuite également des manières de faire nouvelles, dont les idées et présentations de SSB et IPFS sont des exemples. Une même technologie—ici, la liste chaînée—peut donc être pensée différemment selon les cultures où elle se déploie, ce qui va alors pondérer ce qui peut être fait avec cette technologie.

---

[17] Voir l'article mis à disposition ici: https://dominictarr.github.io/secret-handshake-paper/shs.pdf

[18] Voir notamment le commentaire d'Ian Bogost sur le sujet, consultable à https://www.theatlantic.com/technology/archive/2017/05/meet-the-counterantidisintermediationists/527553/

[19] "Dev discussions and tutorials are an essential part of our community, and should not be obscured or downplayed."

Cette étroite connexion entre technologie et culture, cette considération du protocole comme artefact relationnel, se manifeste dans la documentation du protocole. Celle-ci oscille entre rigueur technique et clins d'oeil charmants: présentation de la documentation comme une *carte au trésor*, ou encore la représentation de l'échange de clés cryptographiées comme relation érotique[20].

Enfin, un dernier example se trouve dans la manière dont ces protocoles font vivre leur communauté, notamment en termes d'évènements organisés pour du *community-building*. Pour les deux protocoles, cela se manifeste sous formes de camps. Le *IPFS Camp* consiste à rencontrer des *pionniers*, prêts à *hacker* pendant cinq jours, le tout sponsorisé par des entreprises dont IPFS est le fond de commerce[21]; tandis que, de sont côté, le *Scuttlebutt Camp* est un rassemblement sans véritable fin déterminée, ni agenda particulier[22]. Même si la pandémie de la Covid-19 a été responsable de la non-tenue de ces évènements, nous voyons toujours la différence entre IPFS et SSB en termes de conception de ce qu'est un contenu, un individu, un but et un procédé, au sein d'une similarité organisationnelle.

### 3.3. Applications

Autant SSB et IPFS proposent directement une implémentation des clients (c'est-à-dire de logiciels manifestant les protocoles sous leur forme concrète et active, écrits dans des languages particuliers pour des plateformes particulières), autant les

---

[20] Voir le visuel suivant: https://dev.scuttlebutt.nz/assets/handshake-erotica.png

[21] Voir aussi le site hérbergeant les informations sur le camp: https://camp.ipfs.io/

[22] La version SSB se trouve à l'addresse suivante: https://two.camp.scuttlebutt.nz

Critiques protocolaires d'Internet: comparaison des projets IPFS et SecureScuttleButt

domaines d'application de SSB sont beaucoup plus concentrés que ceux d'IPFS. Tandis qu'IPFS, fidèle à sa vision d'expansion globale, privilégie la quantité relative d'applications possibles (plus de 55 listées sur le site d'IPFS[23]), SSB se concentre sur la qualité, en ne listant qu'une application principale suggérée: un réseau social—c'est à dire un échange de messages entre personnes[24].

Cette visée de l'application est clairement établie dès les premières pages de "comment rejoindre SSB". Sans rentrer dans les détails du protocole, sont proposées directement les différentes applications pour rejoindre le *Scuttleverse*, avec en second plan le genre d'applications qui découlent de l'implémentation d'un protocole techniquement plus large que ses visées sociales (e.g. contrôle de version de code, maintenance et distribution de bibliothèques de code, allant même jusqu'à rejoindre le rôle d'IPFS par le biais de *ssbdrv*[25], un système de fichiers basé sur SSB.).

Particulièrement, la fin de l'introduction à SSB consiste autant en un type différent de protocole: une fois que l'installation et l'inscription sont finies, le discours de SSB mentionne non pas le protocole technique, mais un protocole social, mentionné par le terme *tradition*, celle de se présenter sur le canal *#new-people*, mettant une fois de plus en avant leur volonté de prioriser l'aspect social rendu possible par un protocole, plutôt qu'existant dans une stricte isolation technique.

### 4. L'internet, l'espace et le temps

Bien que les deux protocoles étudiés ici, IPFS et SSB, soient similaires dans leur intention pratique de développer un protocole permettant la distribtion décentralisée et cryptographiée d'information entre pairs d'un réseau par le biais d'un même système technique, la posture théorique des projets se tient néanmoins en porte-à-faux. Nous explorons ici cette tension entre similarité technologique et différences culturelles, entre déterminisme et imagination.

Manuell Castells estimait, dès la fin des années 1990, que la société connectée, société dont les conditions matérielles de réalisation comprennent TCP/IP et HTTP, a fait entrer une grande partie du monde, dans une ère d'espaces de flux et de temps atemporel (*space of flows and timeless time*) (Castells, 2009). L'espace de ces sociétés est un espace qui permet, par des moyens technologiques, de réaliser une simultanéité sans pour autant demander une contiguité, accompagné d'une temporalité qui, tendant à l'immédiateté, tend à s'effacer elle-même.

---

[23] Voir la liste des applications du protocole IPFS: https://docs.ipfs.io/concepts/usage-ideas-examples/

[24] De manière générale, SSB présentait au 01.04.2022 32 applications au total, dont 13 autour du concept de réseau social. Voir la liste des applications du protocole SSB: https://handbook.scuttlebutt.nz/applications

[25] Voir le code source de SSBDRV ici: https://github.com/cn-uofbasel/ssbdrv/

D'après Castells, même si l'espace se transforme, au début des années 1990, du matériel au dématérialisé, l'inter-opérabilité des lieux de décisions de ces centres d'opérations ne change pour autant pas leur statut de *centres*, de noeuds principaux par lesquels doivent transiter les noeuds secondaires afin de communiquer. Cette situation de réseau centralisé est elle-même une conséquence tant de dynamiques économiques et de services marchands, que d'un protocole (HTTP) impliquant la limitation de la duplication de l'information hébergée[26].

C'est cette centralisation qui est le premier objet des critiques des deux protocoles étudiés ici. Ceux-ci nous montrent cependant qu'une telle situation de simultanéité peut, par ces mêmes moyens techniques, être repensée de diverses manières, et notamment en prenant en compte les communautés imaginées en tant qu'utilisatrices de ces protocoles, les priorités discursives des mainteneurs et mainteneuses des protocoles, considérant donc de manière plus holistique la conception d'un protocole, son implémentation et ses applications comme éléments indissociables d'un même discours, au-delà d'une simple spécification technique.

D'une part, IPFS suggère que le réseau peut pousser encore plus loin cette dynamique, se basant sur une interprétation particulière de la vision originelle de Tim Berners-Lee, pour aboutir à un dispositif où le réseau est intégré directement dans chaque poste via un système de fichier. De cette manière sont contournés les côtés négatifs de la centralisation, notamment la disparition de contenus, et la lenteur de téléchargement des données va être résolue par un système d'identification et d'accès à ces contenus.

D'autre part, SSB approche le problème sous l'angle inverse. Il s'agit de présenter une critique pratique de l'impératif d'être connecté en permanence, à travers de larges conglomérats économiques, une nécessité qui obfusque la réalité que la synchronisation "authentique" entre individus est toujours repoussée un peu plus, créant un désir d'immédiateté accru, un phénomène que nous relions aux travaux par Dominic Pettman

---

[26] Excepté pour ce qui est des téléchargements et des caches

Critiques protocolaires d'Internet: comparaison des projets IPFS et SecureScuttleButt

sous le nom de *will-to-synchronize* (Pettman, 2015). SSB propose une vision du monde où l'espace et le temps de chacun des membres du réseau n'est pas nécéssairement identique, et où la proximité physique est une condition suffisante à l'échange d'information.

Des positions similaires sur la centralisation, mais opposées sur la mise-à-disposition, donc. IPFS recherche une quasi-universalité (ou, selon leurs termes, une inter-planétarité), alors que SSB recherche l'implémentation technique et culturelle d'une priorité au local. Et pourtant, ces deux protocoles, ces deux visions divergentes se basent sur des algorithmes extrêmement similaires, nous permettant alors de nous poser la question du déterminisme technologique.

### 4.1 Déterminismes socio-technologiques

Dans les deux cas, nous avons à faire à des protocoles basés sur le principe de l'*append-only log*, c'est à dire une suite d'entités qui ne peut que s'accroître, et dont les entités précédentes dans la liste sont immuables. Ces listes immuables le sont rendues par l'utilisation de techniques cryptographiques utilisant le concept de liste chaînée. Et pourtant, un protocole est un objet socio-technique qui s'applique également à un *problem domain*, le domaine d'application de l'algorithme au-delà de son aspect strictement technologique et computationnel, et c'est ce domaine d'application qui emmène la critique protocolaire au-delà du strictement technique. La principale différence n'est donc pas la technologie mais le choix du domaine d'application: IPFS applique son protocole au contenu, possédant une relation à d'autres contenus (fichiers ou dossiers), et une relation à soi-même (établissant par la-même l'immutabilité, et la pérennité du contenu en question), disséminant cette information à travers des mécanismes de marché financier. À l'opposé, SSB utilise ces outils techniques pour définir l'individu, chaque noeud du réseau, comme propriétaire des l'information—socialisée, c'est-à-dire une chaine d'informations basées sur un graphe social, d'individus concrets qui disséminent l'information par proximité physique, ou par proximité sociale via les *pubs*.

La communauté d'IPFS, lorsqu'il s'agit de développer des applications pour le protocole, se retrouve toujours à considérer une approche globale, centralisée, pour un moteur de recherche, ou encore à s'enquérir d'une architecture de réseau social qui semble être trop compliqué pour être véritablement centralisé. Particulièrement révélateur est un des deux exemples donnés par

Pierre Depaz

les membres d'IPFS comme exemple de la résilience du protocole face à la censure: lors du référendum catalan de 2017, le site pour s'enregistrer en tant que que votant ou votante avait été censuré par les autorités de Madrid, et le site avait été mis en ligne sur IPFS en tant qu'alternative permanente d'un contenu victime de censure; pourtant, comme le note l'utilisateur Akira:

> Malheureusement, la plupart des utilisateurs de l'époque utilisaient la passerelle HTTP gateway.ipfs.io 6, qui était également censurée, mais les utilisateurs plus avisés évitaient la censure en utilisant un démon IPFS normal et en installant la simple extension de navigateur IPFS Companion. IPFS est désormais beaucoup plus facile à utiliser.[27][28]

La technologie était donc là, mais l'usage ne semble pas avoir suivi. De manière plus générale, l'écosystème d'IPFS semble se focaliser plus sur l'existence d'applications, que sur leur usage, tel que le montre également la page *Awesome IPFS*, dont environ la moitié des applications sont désormais désuettes ou non-maintenues[29] sur la centaine de disponibles.

Plus qu'un déterminisme technologique, SSB présente une approche mutuellement informative entre technique et culture, imprégnée de dynamiques holistiques à travers laquelle protocoles et humains co-existent et co-agissent dans un seul et même système. Tel que le présente la documentation du protocole, ce dernier est efficace non pas strictement par ses vertus techniques mais par l'environnement interpersonnel que procure la communauté utilisant ce protocole:

---

[27] Unfortunately, most users back then used the HTTP gateway gateway.ipfs.io 6, which was also censored, but tech-savvier users avoided censorship using a regular IPFS daemon and installing the simple IPFS Companion browser extension. IPFS is now way easier to use.

[28] Voir https://discuss.ipfs.io/t/how-censorship-resistant-is-ipfs-intended-to-be/7892/5 pour cette discussion des limites sociales d'une technologie anti-censure.

[29] Consulté le 21.02.2022, à l'addresse https://awesome.ipfs.io/apps/

Critiques protocolaires d'Internet: comparaison des projets IPFS et SecureScuttleButt

> *L'une de ses premières applications a été le réseau social, et il est également devenu l'un des plus attrayants, car les personnes qui s'y retrouvent ne sont pas des imbéciles.*[30] *(Anonymous, 2021)*

Cet enchevêtrement entre développements technologiques et interactions sociales est présente dès la genèse de SSB, racontée par Dominic Tarr: il développe des concepts technologiques, les présente et intègre des contributions d'autres individus et collaborateurs, puis développe le protocole d'avantage, dans un mouvement de balance qui contraste avec l'aspect plus téléologique d'IPFS (Tarr, 2016).

## 5. Conclusion

Cette étude comparative de deux protocoles nous a donc permis présenter le concept de critique protocolaire: une critique tant technique que discursive et opérationelle. Dans les deux cas d'étude, nous avons pu voir que la constitution d'une critique protocolaire trouve son point de départ dans une description technique, puis d'une pluralité de discours soutenant et justifiant une une manière de communiquer réifiée dans la technique, et enfin confrontée à une application pratique.

Cela nous a également permis de révéler la manière dont une critique protocolaire, notamment à travers le dispositif technique, présente des présupposés profonds quant à la manière dont la communication d'une information doit se dérouler, lorsqu'il est appliqué à un domaine en particulier, que ce soit en se focalisant sur le message (IPFS) ou sur les interlocuteurs ou interlocutrices (SSB).

D'autre part, nous avons pu élucider comment cette rhétorique du protocole peut plus ou moins s'aligner à la rhétorique des utilisateurs (notamment en ce qui concerne la censure sur IPFS, ou en ce qui concerne la centralisation des données, ou l'application à des domaines plus spécifiques), et identifier certaines cohérences et incohérences entre les discours autour du protocole et les applications développées par ce protocole, touchant à la question du déterminisme technologique, que ce soit sur une trajectoire téléologique (IPFS), ou sur une trajectoire plus écologique (SSB). En fin de compte, donc, un protocole de communication semble impliquer un certain déterminisme par sa technologie, mais aussi par sa culture.

---

[30] One of its first applications was as a social network, and it has also become one of the most compelling because the people who hang out there are not jerks.

Pierre Depaz

**Références**

Critiques protocolaires d'Internet: comparaison des projets IPFS et SecureScuttleButt

Pierre Depaz

Critiques protocolaires d'Internet: comparaison des projets IPFS et SecureScuttleButt